# A Review of Man-in-the-Middle Attacks

Subodh Gangan

*Abstract:* This paper presents a survey of man-in-the-middle (MIM) attacks in communication networks and methods of protection against them. In real time communication, the attack can in many situations be discovered by the use of timing information. The most common attacks occur due to Address Resolution Protocol (ARP) cache poisoning, DNS spoofing, session hijacking, and SSL hijacking.

**Introduction**
Man-in-the-Middle (MIM) attacks make the task of keeping data secure and private particularly challenging since attacks can be mounted from remote computers with fake addresses. Whereas communications security was primarily that of breaking of encryption transformations (as in the case of the Enigma machine during the Second World War [1]), the problem of security in computer networks also involves active interference by intruders, and of these MIM attack is one of the most spectacular. The MIM attack takes advantage of the weaknesses in the authentication protocols being used by the communicating parties. Since authentication is generally provided by third parties who issue certificates, the system of certificate generation becomes another source of potential weakness.

The MIM attack allows the intruder or the unauthorized party to snoop on data through the backdoor. This intervention is also being used by companies to pry upon their employees and for adware. For example, in early 2015, it was discovered that Lenovo computers came preinstalled with adware called Superfish that injects advertising on browsers such as Google Chrome and Internet Explorer. Superfish installs a self-generated root certificate into the Windows certificate store and then resigns all SSL certificates presented by HTTPS sites with its own certificate. This could allow hackers to potentially steal sensitive data like banking credentials or to spy on the users' activities [2].

Cryptographic protocols designed to provide communications security over a computer network are a part of Transport Layer Security (TLS). These protocols use X.509 which is an ITU-T standard that specifies standard formats for public key certificates, certificate revocation lists, attribute certificates, and a certification path validation algorithm [3]. The X.509 certificates are used for authentication the counterparty and to negotiate a symmetric key. As mentioned, certificate authorities are a weak link within the security system. In electronic mail, although servers do require SSL encryption, contents are processed and stored in plaintext on the servers.

Google and Mozilla recently announced that they would stop accepting certificates issue by CNNIC (China Internet Network Information Center) [4]. According to Google's Security Blog [5]:
> On Friday, March 20th, we became aware of unauthorized digital certificates for several Google domains. The certificates were issued by an intermediate certificate authority apparently held by a company called MCS Holdings. This intermediate certificate was issued by CNNIC. CNNIC is included in all major root stores and so the misissued certificates would be trusted by almost all browsers and operating systems. …
> We promptly alerted CNNIC and other major browsers about the incident, and we blocked the MCS Holdings certificate in Chrome with a CRLSet push. CNNIC responded on the 22nd to explain that they had contracted with MCS Holdings on the basis that MCS would only issue certificates for domains that they had registered. However, rather than keep the private key in a suitable HSM, MCS installed it in a man-in-the-middle proxy. These devices intercept secure connections by masquerading as the intended destination and are sometimes used by companies to intercept their employees' secure traffic for monitoring or legal reasons. The employees' computers normally have to be configured to trust a proxy for it to be able to do this. However, in this case, the presumed proxy was given the full authority of a public CA, which is a serious breach of the CA system…This situation is similar to a failure by ANSSI in 2013.

Consider two individuals, namely Alice and Bob, communicating with each other. The third-party, Eve, is the MIM attacker. The communication between Alice and Bob starts after the initial handshake to authenticate each other. But if the authentication protocol is not strong then Eve can impersonate as the other communicating party to both Alice and Bob.

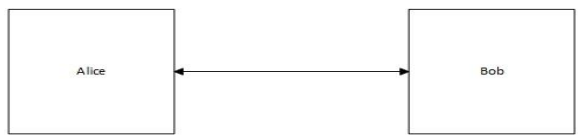

Figure 1: Normal communication between Alice and Bob

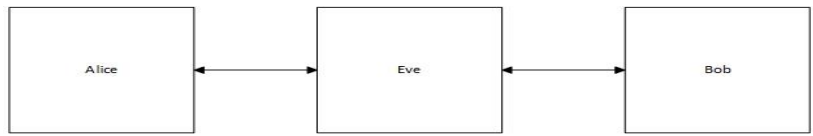

Figure 2: MIM attack performed by Eve

In the MIM attack, Eve will start its communication with both Alice and Bob. The messages between Alice and Bob will go through Eve as Eve will impersonate Bob as Alice and Alice as Bob. Both Alice and Bob won't know about Eve intercepting their messages. Therefore, Eve take advantages of Alice and Bob without their consent and obtains vital information.

The hallmark of the knowledge society is uncertainty [6],[7] together with the problems of privacy and identity in the networked world [8]-[9]. Traditional techniques of cryptography [10] that use random sequences are one way to deal with these issues at the level of signal integrity [11]-[15]. Backdoor entry, of which MIM is one, can be used by good and bad actors [16]-[18]. One way to avoid certain MIM attacks is the implementation of the double lock protocol but it was observed that this method was not successful because the man in the middle attack was still persistent inside the network [19].

The Double Signature Double Lock Protocol protocol uses hashing technique where the signatures of the 2 parties communicating are exchanged with the help of hashed digits and the signatures are verified. If the signatures don't match then the communication doesn't take place and the man in the middle attack can be avoided because the attacker would be completely unaware of the hashed digit used for authentication of the communication. This paper presents a survey of the problems with authentication that can be exploited to mount MIM attacks.

**Examples of MIM attacks**
An example of an offline MIM attack is the interception of a letter by the mailman who either just reads its contents or even replaces its contents. One can visualize an online MIM attack in a public place like a mall that provides free Wi-Fi connection available that has a wireless router with malicious software installed in it. If a user visits a bank's website at that time from phone or laptop, he or she may end up losing bank credentials. These attacks can be caused because of the following reasons:

1. ARP Cache Poisoning
2. DNS Spoofing
3. Session Hijacking
4. SSL Hijacking

**Address Resolution Protocol Communication (ARP)**
In the normal ARP communication, the host PC will send a packet which has the source and destination IP address inside the packet and will broadcast it to all the devices connected to the network. The device which has the target IP address will only send the ARP reply with its MAC address in it and then communication takes place.  The ARP protocol is not a secured protocol and the ARP cache doesn't have a foolproof mechanism which results in a big problem.

The ARP reply packet can be easily spoofed and it can be sent to the machine which sent the ARP request without knowing that this is not the actual machine, but an attack to cause data breaches. This happens because the ARP cache table will be updated as decided by the attacker and so all the network traffic will go through the attacker and he will have all the data and make the most of it. This is the best kind of an attack on the Local Area Network.

Different types of tools are available in the market for ARP Cache poisoning some of them are Ettercap, Dsniff and Cain and Abel's etc. We can try to control this ARP cache poisoning by using Dynamic ARP Inspection (DAI). DAI is a security feature that is used to validate the ARP packets in a network and to discard the invalid IP to MAC address bindings. We need to carry out this inspection on the Ethernet switches by manually configuring them, but we can't do these on the switches which aren't compatible with this inspection [20].

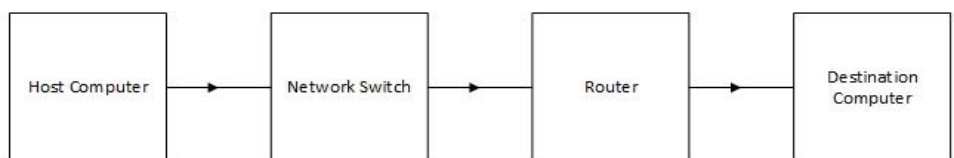
Figure 3: Regular traffic flow between two Computers

The ARP request and replies don't require any authentication or verification since all the hosts on the network will trust the ARP replies. We can also update the ARP cache table by inserting a static entry for the gateway so the attacker wouldn't manipulate with the gateway entry in the table but this is not a fool-proof solution as we need to change the ARP table gateway entry regularly whenever we relocate. The secure socket layer is used by HTTP or transport layer for secure connection. The web browser will look for certificates of the web server and authenticate its validity. If the certificate is authenticated we will have a secure connection but if we have any problem with the certificate it would say as untrusted.

**ARP Cache Poisoning**
In ARP cache poisoning, the attacker would be sniffing onto the network by controlling the network switch to monitor the network traffic and spoof the ARP packets between the host and the destination PC and perform the MIM attack.

**DNS Spoofing**
The target, in this case, will be provided with fake information which would lead to loss of credentials. As explained earlier this is a kind of online MIM attack where the attacker has created a fake website of your bank, so when you visit your bank website you will be redirected to the website created by the attacker and then the attacker will gain all your credentials. Whenever we enter a website on our PC, DNS request is sent to the DNS Server and we will get a DNS reply message in the response. This is shown with the help of figure 4.

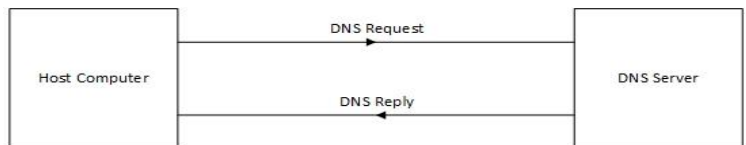
Figure 4: DNS Communication between the host and the DNS Server

This DNS request and reply are mapped together with a unique identification number. If the attacker gets hold of the unique identification number then by disguising the victim with a corrupt packet containing the identification number the attack can be launched. The attacker redirects the victim to the fake website by performing ARP cache poisoning to detour the DNS request message to which the fake reply packet is sent.

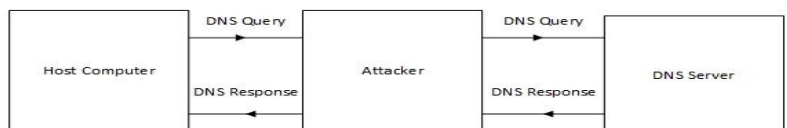
Figure 5: The attacker performs MIM attack using DNS Spoofing

The host computer wants to connect to a website so it will send a DNS query request to the DNS server but due to the MIM attack the attacker will intercept this DNS query and send a fake DNS reply to the host PC. The host PC wouldn't come to know whether the response is legitimate or not and it will start communicating with the malicious website of the attacker which causes data breaches.

**Session Hijacking**
As the name suggests a session is established when we have a connection between the client and a server. Transmission Control Protocol is referred as a session since it first establishes a connection, then transfers the data and finally terminates the connection. This is known as the 3-way handshake process which shows how a proper session looks like. One of the popular session hijackings is done by stealing cookies with the help of Hyper Text Transfer Protocol (HTTP).

In any website one requires username and password for authentication and establishing the session. Once the session is established, unless and until one logs out the session is not terminated, so to administer session cookies are used that provide information that the session is still continuing. If the attacker gets hold of this cookie he/she can have the session information which might be secretive.

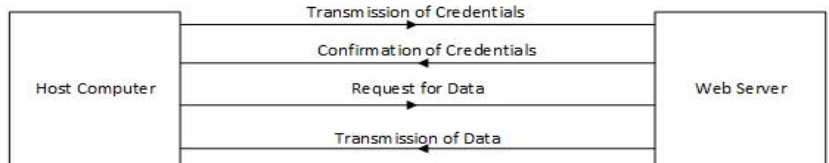
Figure 6: Session Establishment between the host PC and Web Server

The figure shows how a normal session is established between the host PC and the Web Server. As one can see the session cannot be established without authentication.

**Session hijacking attack**
Once the session is established between the host PC and the web server the attacker can obtain certain parts of the session establishment which is done by capturing the cookies that were used for the session establishment. As shown in the figure 6 after the session establishment the attacker obtains the session information between the host PC and the web server and controls the session.

**Secure Socket Layers**
One needs to provide security while communicating with network devices which can be obtained with the help of Secure Socket Layers (SSL) or Transport Layer Security (TLS) by using encryption methodology. One uses this protocol with other protocols for secure implementation of the services that the protocol provides. HTTPS is the most commonly used protocol and most of the online banking services and email services use it to ensure security between their servers and your web browser.

In order to understand how exactly this protocol works consider the following example. Suppose the host PC wants to connect to yahoo mail account then the communication process starts as stated below:
1. Using HTTP port 80, the client web browser will connect to http://mail.yahoo.com.
2. The web server will process this request and redirects the client to the HTTP version of this website using HTTP code 302 redirect.
3. The client will now connect to https://mail.yahoo.com using port 443.
4. The server will provide a certificate to the host PC to verify the identity of the website using the digital signature.
5. The host PC will now verify this certificate with the list of certificates it has.

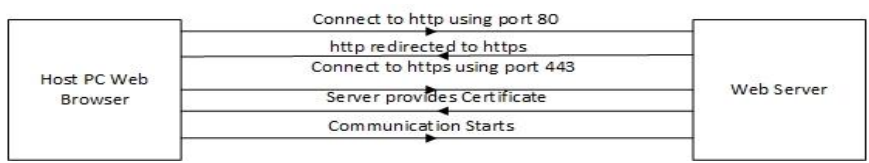

Figure 7: SSL protocol communication

If the certificate doesn't match with the list of certificates of the host PC then we say that the website has failed to verify its identity so the host PC will get a certificate validation error. Even after we get this error we can proceed to connect to the website but it might be risky because we won't know whether it is the actual website we need to connect to. This SSL attack is explained more precisely with the help of a practical example.

**Current Scenario**
As we all know that there are lots of people surfing the internet and the numbers are increasing day by day. So it becomes necessary to protect the data that is available on the internet and maintain internet security because there are lot many bad guys out in the world who would like to take advantage of this and have data breaches and different kinds of attacks without being get noticed.

As mentioned in the Introduction, CNNIC, which is dministered by Cyberspace Administration of China (CAC), issued some untrusted digital certificates called as the test certificates, to the different companies and the general public, which were valid for two weeks and would get expired by April 2015 [21].

By loading this certificate into the firewall device, the firewall would issue certificates for different domains and an SSL MIM attack could be performed [21]. This is because almost each and every operating system and web browsers will trust this altered certificate and it would cause a major problem in the Internet Certificate Authority system.

By trusting the certificate the web browser would create problem for itself since the certificate holder could easily decode and observe the communication going on inside the network without providing any prior warnings to the web browser. As discussed earlier the bad guys present outside would have their fake website to appear whenever the users would like to visit a specific website and without knowing the users would be opened to the MIM attack.

The web browsers update their current version in order to take into account these loopholes and release them into the market. The user will then upgrade to this latest version and remain updated. So what we have found out is that the today's internet still relies on to encrypt the sensitive data. Since almost all the operating systems trust most of the certificates it takes just one fake certificate to take advantage of this situation by compromising the security of the complete system.

Another example of the MIM attack is the SSL attack for popular domains such as google.com, yahoo.com, live.com, and skype.com [21]. The fake SSL certificates were issued by Comodo, the trusted certificate authority. But the attack was discovered and the false certificates were revoked.

The web services would be altered because of this certificate facilitating different kinds of attacks such as phishing attacks. In order to check the validity of TLS certificates, the browsers use revocation lists. But these revocation lists are unavailable if the internet has outage. It has therefore been suggested that the untrusted certificates need to be permanently coded into the browsers. Most of the operating systems available now have an automatic updater installed for the revoked certificates. The automatic update will automatically download the canceled certificate without the knowledge of the user.

Figure 8 shows us the month wise statistical data from December 2013 to November 2014 by Symantec Corporation of the different browsers used in our day to day life. We can depict from the figure that the Internet Explorer is the most vulnerable to the diversified attacks.

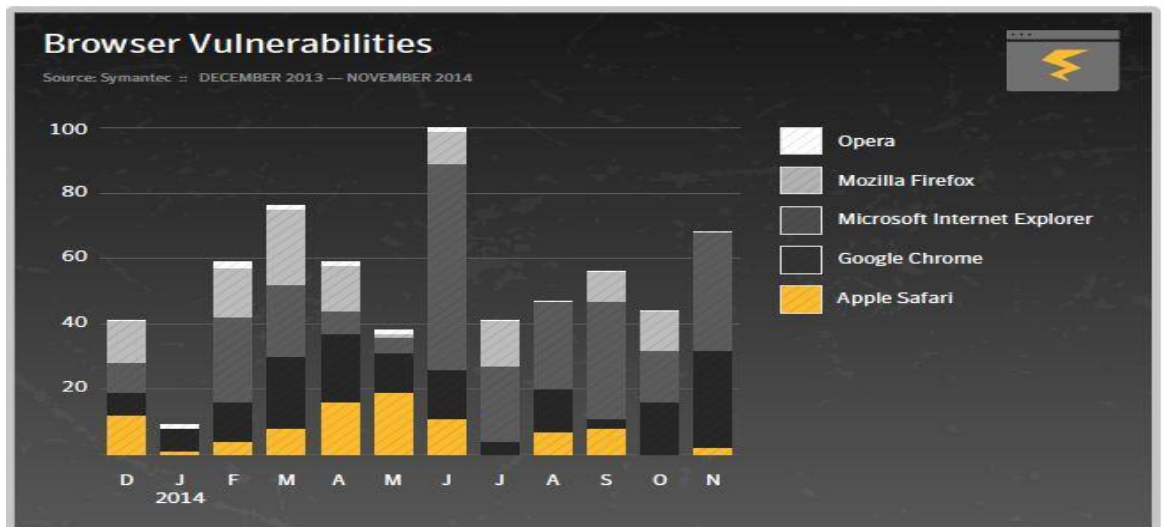

Figure 8: Graphical Representation of the different web browsers [22]

As we can see from Figure 9, the majority of the network attacks are Denial of Service, Brute force and browser attacks. The Denial of Service, browser and SSL attack all constitute the MIM attack.

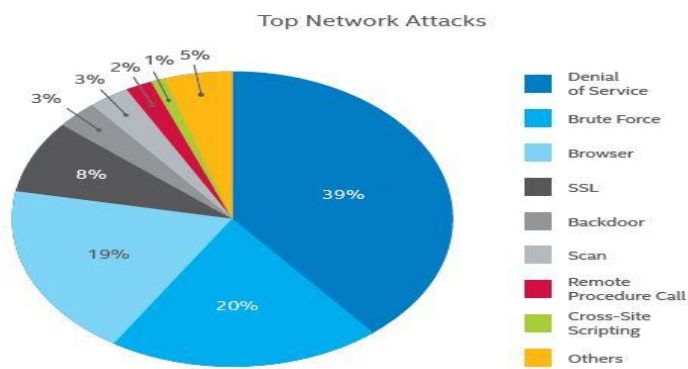

Figure 9. Top Network Attacks (McAfee Labs [22])

**Practical Implementation**

In order to understand the MIM attacks in a better way, we discuss a small laboratory experiment. The attack is based on DNS spoofing and SSL stripping, the operating system used for performing the attack is Kali Linux (64 bit). This operating system has some inbuilt tools for the network sniffing, ARP poisoning, online attacks and network scanners etc.

In the experiment, carried out virtually from home, connecting to the VPN of the Tulsa IA Lab so that I can have a number of live hosts on the network to perform an attack. The Cisco AnyConnect Secure Mobility Client VPN software was used to connect to the lab virtually. The Tulsa IA Lab has ESXI Servers to which we get connected remotely with the help of this software and VSphere Client, hence we get connected to the network.

**DNS Spoofing Attack**

1: The ettercap tool which we use to perform the MIM attack has an inbuilt file etter.dns which we need to modify so we need to find the location of the file.
    Syntax: locate etter.dns

2: We will get the location of the file from the previous step and then we need to edit this file as per our convenience to target the victim.
    Syntax: vim /etc/ettercap/etter.dns

3: We want the user to get denied to access facebook.com so we just add a line in the etter.dns file i.e. we enter a host name and create a pointer for it. In this example, the pointer is pointing to my IP address of Kali Linux (10.15.0.9) but we can create any pointer as per our convenience. Since I already knew my IP address I just added the pointer, otherwise I would have to use the command ifconfig to find my IP address. Whenever we make changes make sure we write and quit before exiting.
    Syntax: *.facebook.com A 10.15.0.9

4: We will start a web server on my Kali Linux so that whenever someone tries to open a specific domain he/she will be directed here.
    Syntax: service apache2 start

5: We can also create an index file for our web server which the victim will see whenever he/she tries to access the web page that we have targeted.
    Syntax: cat /var/www/index.html

6: We need to figure out our gateway so that the ARP Poisoning attack can be performed and redirect all the requests to our Computer instead of the gateway.
    Syntax: netstat –nr (To find the gateway)

    7: DNS Spoof plugin is used to perform this attack.
    Syntax: ettercap –Q –T –M arp –P dns_spoof /10.15.0.254/ /10.15.0.7/
As you can see in this syntax we are specifically targeting a victim whose IP is 10.15.0.7 but we can also target the whole network just by excluding the target machines IP.

8: The DNS spoof plug in will activate now and the victim will be spoofed. Now as soon as the victim will open facebook.com in his web browser he will be redirected to the web server created by us.

9: As you can see in the image below that the victim tried to visit the facebook.com and the permission is denied which means we are successful in performing the attack. You can also check in the address bar which says it's not a secured connection. So whenever this kind of attacks takes place you should verify whether it's a secure connection or not.

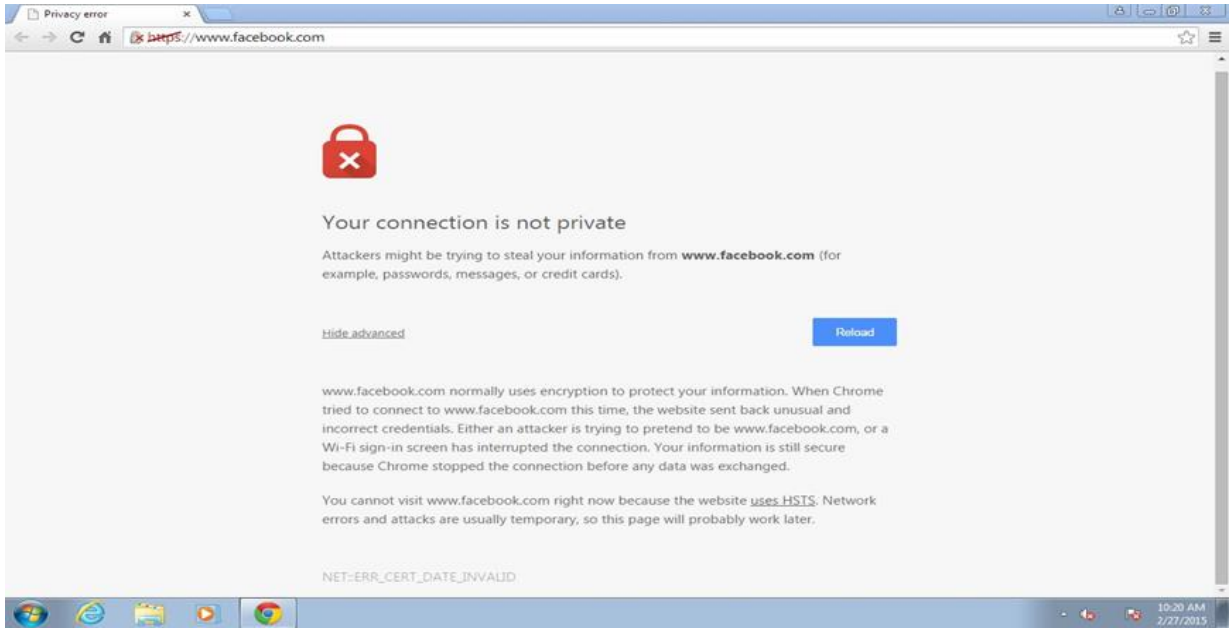
Figure 10: Victim denied access to the website

**ARP Spoofing and SSL Stripping**
The block diagram showed bellow tells us about what happens exactly when we perform these kinds of attacks.

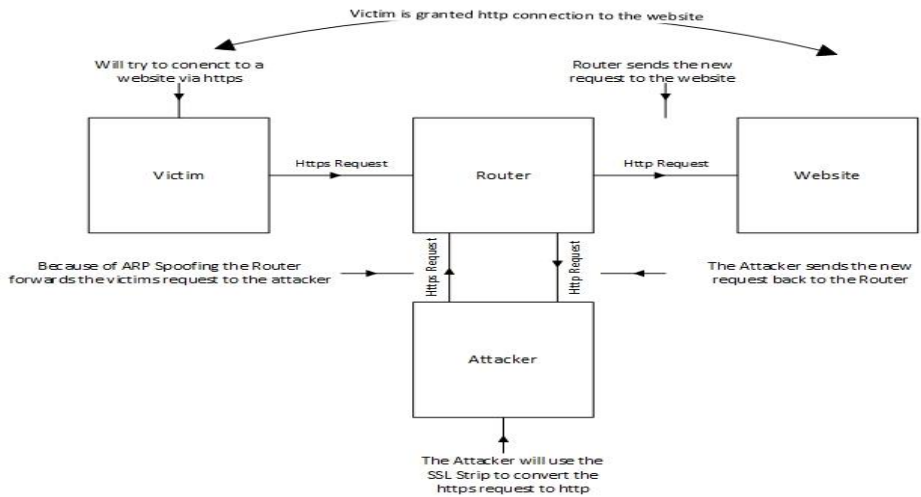
Figure 11: Practical representation of the devices functionality

1: In order to establish this attack we need to enable IP forwarding. We need to check the value to verify whether IP forwarding is pre-configured or not. If we get the value as '1' then we say that IP forwarding is enabled but if we get the value as '0' then we need to enable the IP forwarding.

Syntax: echo 1 > /proc/sys/net/ipv4/ip_forward     (To enable IP forwarding)
Syntax: cat /proc/sys/net/ipv4/ip_forward           (To check the value)

2: We need to redirect the traffic from source port 80 to the destination port 8080 and the traffic should mainly consist of TCP. All of this will take place by updating the NAT table. Whenever the conditions would match the packets will be redirected to port 8080 as specified.
Syntax: iptables –t nat –A PREROUTING –p tcp –destination-port 80 –j REDIRECT –to-port 8080

3: Here also we need to find the IP address of the gateway. This is similar to the step 6 of the previous example.

4: In order to identify any live host on the network we can use the Nmap software or else if we know the IP address of the target machine we can directly use that IP address. In our case, we already know our target and the IP address of the target is 10.15.0.15

5: Before performing the SSL stripping we need to redirect the HTTP traffic. This is done with the help of ARP Spoofing.
Syntax: arpspoof –i eth0 –t 10.15.0.15 –r 10.15.0.254

6: In order to obtain the credentials for a particular website we use SSL stripping. Suppose the user visits www.facebook.com, in order to login to facebook the user needs to login with his credentials and by performing SSL stripping we can steal these credentials (user id and password).
Syntax: sslstrip –l 8080

7: Once the SSL strip is established we can get the credentials in a log file which can be accessed easily. We can see the image below which shows us that we were successful in performing the attack and stealing the credentials.
Syntax: cat sslstrip.log

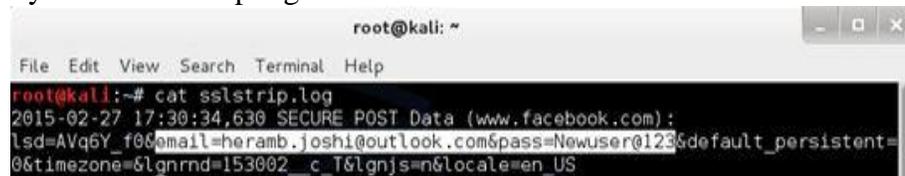

Figure 12: Successful attacked performed to obtain the user credentials

So in order to avoid ARP cache poisoning and DNS spoofing practically we need to monitor the network thoroughly to identify the probe packets coming inside the network for performing the attack. We need to build a strong and secure network with the help of firewalls, Intrusion Detection, and Prevention Systems. If we are using the web browser to login to a particular website with our credentials we need to make sure that we log in securely. This can be verified with the syntax HTTPS instead of HTTP on the URL tab. What HTTP will do is that it will provide a secure connection between you and the website used so that we would not face any breach.

**Concluding Remarks**

Two general points: (i) The MIM attack also applies to quantum cryptographic system [23]-[27]. Indeed, that is the easiest way to break a quantum cryptographic system if there are not enough safeguards in the authentication of the two parties during their communication [28]; (ii) such attacks are also important in P2P networks [29]. In a real-time communication scenario the MIM attacker can be found out by consideration of timing information.

ARP poisoning can be avoided by using the shell script running at the backend which will keep a track of entries in the ARP cache table (maps IP address and MAC address) [30],[31]. The problem with this solution is that because of periodically generated ARP request there is a lot of traffic inside the network and the shell script will run only on Linux, not on windows.

Since we cannot eliminate the MIM attack completely we can try to minimize the possibilities of this attack onto the network. Some of these security measures include host hardening i.e. operating systems onto the network should be upgraded, network designing from security point of view, network devices and the computers onto the network should be updated periodically and the patches should be installed regularly. In this way by trying these options practically we might try to ensure that the MIM attacks could cause less harm to us. Nevertheless, MIM attack will remain an attack of choice not only for bad actors but also for surveillance groups [32].

**References**
1. H. Sebag-Montfiore, Enigma, The Battle for the Code. Weidenfield & Nicolson, 2000.
2. I. Paul, Lenovo preinstalls man-in-the-middle adware that hijacks HTTPS traffic on new PCs. PC World, Feb 19, 2015.
3. D. Taylor, Ed. RFC 5054: Using the Secure Remote Password (SRP) Protocol for TLS Authentication. Internet Engineering Task Force. November 2007.
4. A. Henochowicz, Minitrue: Man-in-the-middle Attacks Enabled by CNNIC, China Digital Times, 2015.
5. A. Langley, Maintaining digital certificate security. Google Online Security Blog, March 23, 2015.
6. S. Kak, The Nature of Physical Reality. Peter Lang, New York, 1986, 2011.
7. S. Kak, The Architecture of Knowledge. CSC, New Delhi, 2004.
8. B. Wellman, The rise (and possible fall) of networked individualism. Connections, 2002.
9. B. Wellman, Computer networks as social networks. Science 293: 2031-2034, 2001.
10. S. Singh, The Code Book: the Secret History of Codes and Code-breaking. Fourth Estate, London, 1999.
11. S. Kak, Classification of random binary sequences using Walsh-Fourier analysis. IEEE Trans. on Electromagnetic Compatibility, EMC-13: 74-77, 1971.
12. S. Kak and A. Chatterjee, On decimal sequences. IEEE Trans. on Information Theory IT-27: 647 – 652, 1981.
13. S. Kak, Encryption and error-correction coding using D sequences. IEEE Trans. on Computers C-34: 803-809, 1985.
14. D. Eastlake 3rd, S. Crocker, J. Schiller, Randomness Recommendations for Security. Network Working Group, MIT, 1994.
15. J. Yan, Password memorability and security: Empirical results. IEEE Security and Privacy, 2004.

16. Philip R. Zimmermann. The Official PGP User's Guide. MIT Press, Cambridge, MA, USA, 1995.
17. https://www.gnupg.org/ GNU Privacy Guard/
18. S. Heinrich, Public Key Infrastructure based on Authentication of Media Attestments. arXiv:1311.7182
19. Seung Yeob Nam, Dongwon Kim and Jeongeun Kim, Enhanced ARP: Preventing ARP Poisoning-Based Man-In-The-Middle Attacks, Communication Letters 14: 187-189, 2010.
20. Dan Goodin, Bogus SSL certificate for Windows Live could allow man-in-the-middle hacks, Arstechnica, 2015.
21. Ben Nahorney, Symantec Intelligence Report, 2014.
22. McAfee Labs Threats Report (2014).
23. S. Kak, A three-stage quantum cryptography protocol. Foundations of Physics Letters 19: 293-296, 2006.
24. S. Kak, Quantum information and entropy. Int. Journal of Theo. Phys. 46: 860-876, 2007.
25. S. Kak, The piggy bank cryptographic trope. Infocommunications Journal 6: 22-25, 2014.
26. S. Kak, Authentication Using Piggy Bank Approach to Secure Double-Lock Cryptography. arXiv:1411.3645
27. K.W.C. Chan et al., Multi-Photon Quantum Key Distribution Based on Double-Lock Encryption. arXiv:1503.05793
28. L. Lydersen et al., Hacking commercial quantum cryptography systems by tailored bright illumination. Nat. Photonics 4, 686, 2010.
29. L. Washbourne, A Survey of P2P Network Security. arXiv:1504.01358
30. G. N. Nayak and S. G. Samaddar, Different flavors of man-in-the-middle attack, consequences and feasible solutions, Chengdu, 3rd IEEE International Conference Volume 5, pp 491-495, 2010.
31. B. Isaac, Secure ARP and Secure DHCP Protocols toMitigate Security Attacks. Int. Journal of Network Security 8: 107-118, 2009.
32. E. Moyer, NSA disguised itself as Google to spy, say reports. CNET, September 12, 2013.